\documentclass[twocolumn,prb,showpacs,aps,superscriptaddress, reprint]{revtex4-1}

\usepackage[english]{babel}
\usepackage[ansinew]{inputenc}
\usepackage{times}
\usepackage{graphicx}
\usepackage{graphics}
\usepackage{amsmath}
\usepackage{amsfonts}
\usepackage{amssymb}
\usepackage{dsfont}
\usepackage{epstopdf}
\usepackage{makeidx}
\usepackage{subfigure}
\usepackage{color}
\usepackage{pgf}
\usepackage{bm}
\usepackage{tikz} \usepackage[normalem]{ulem}

\newcommand{\be}{\begin{equation}}
\newcommand{\ee}{\end{equation}}
\newcommand{\bea}{\begin{eqnarray}}
\newcommand{\eea}{\end{eqnarray}}

\newcommand\redsout{\bgroup\markoverwith{\textcolor{red}{\rule[0.5ex]{4pt}{0.8pt}}}\ULon}

\makeindex
\begin{document}
\title{Active magneto-optical control of spontaneous emission in graphene}

\author{W. J. M. Kort-Kamp}
\affiliation{Theoretical Division and Center for Nonlinear Studies, Los Alamos
National
Laboratory, Los Alamos, New Mexico 87545, United States}
\affiliation{Instituto de F\'{\i}sica, Universidade Federal do Rio de Janeiro,
Caixa Postal 68528, Rio de Janeiro 21941-972, RJ, Brazil}
\author{B. Amorim}
\affiliation{Instituto de Ciencia de Materiales de Madrid,
CSIC, Cantoblanco E28049 Madrid, Spain}
\affiliation{Department of Physics and Center of Physics,
University of Minho, P-4710-057, Braga, Portugal}
\author{G. Bastos}
\affiliation{Instituto de F\'{\i}sica, Universidade Federal do Rio de Janeiro,
Caixa Postal 68528, Rio de Janeiro 21941-972, RJ, Brazil}
\author{F. A. Pinheiro}
\affiliation{Instituto de F\'{\i}sica, Universidade Federal do Rio de Janeiro,
Caixa Postal 68528, Rio de Janeiro 21941-972, RJ, Brazil}
\affiliation{Optoelectronics Research Centre and Centre for Photonic Metamaterials,
University of Southampton, Highfield, Southampton SO17 1BJ, United Kingdom}
\author{F. S. S. Rosa}
\affiliation{Instituto de F\'{\i}sica, Universidade Federal do Rio de Janeiro,
Caixa Postal 68528, Rio de Janeiro 21941-972, RJ, Brazil}
\author{N. M. R. Peres}
\affiliation{Department of Physics and Center of Physics,
University of Minho, P-4710-057, Braga, Portugal}
\author{C. Farina}
\affiliation{Instituto de F\'{\i}sica, Universidade Federal do Rio de Janeiro,
Caixa Postal 68528, Rio de Janeiro 21941-972, RJ, Brazil}
\date{\today}

\begin{abstract}
We investigate the spontaneous emission rate of a two-level quantum emitter near
a graphene-coated substrate under the influence of an external
magnetic field or strain induced pseudo-magnetic field. We demonstrate that the application of the magnetic field 
can substantially increase or decrease the decay rate.  We show that a suppression as large as 99$\%$ in
the Purcell  factor is achieved even for moderate magnetic fields.  
The emitter's lifetime is a discontinuous function of $|{\bf B}|$, which is a
direct consequence of the occurrence of discrete Landau levels in graphene.
We demonstrate that, in the near-field regime, the magnetic field  enables an unprecedented control of the decay pathways into which the
photon/polariton can be emitted. Our findings strongly suggest that a magnetic
field could act as an efficient agent for on-demand, active control of
light-matter interactions in graphene at the quantum level.
\end{abstract}
\maketitle
The possibility of tailoring light-matter interactions at a
quantum level has been a sought-after goal in optics since the pioneer work of
Purcell~\cite{Purcell-46}, where it was first shown that the environment can
strongly modify the
spontaneous emission (SE) rate of a quantum emitter. To achieve such objective, several approaches
have been proposed so far. One of them is to investigate SE in different system
geometries~\cite{Klimov-Letokhov-1999,Tomas-2001,klimovreview,blanco2004,
thomas2004,carminati2006,RosaEtAl-08,vladimirova2012,Greffet2011,AlvesEtAl2000}.
Advances in
nanofabrication techniques have not only allowed the increase of the
spectroscopic resolution of molecules in complex
environments~\cite{betzig1993}, but have also led to the use of nanometric
objects, such as antennas and tips, to modify the lifetime, and enhance the
fluorescence of single
molecules~\cite{bian1995,sanchez1999,greffet2005,muhlschlegel2005}. The presence
of metamaterials may also strongly affect quantum emitters' radiative
processes. For instance, the impact of negative refraction and
of the hyperbolic dispersion on the SE
have been investigated~\cite{klimov2002, jacob2012,hyperbolicreview}. Also, the influence of cloaking devices on the SE of  atoms has been recently  
addressed ~\cite{kortkamp2013}.

Progress  in plasmonics has also allowed for a unprecedented control
of light-matter interactions at a quantum level. When the emitter is located
near a plasmonic structure it may experience a strong enhancement of the local
field. This effect can be exploited in the development of important applications
in nanoplasmonics~\cite{jackson2004,wei2008,li2010,  klimov2012, Smith2015}. However,  structures
made of noble metals are hardly tunable, which unavoidably
limit their application in photonic devices.  To circumvent these limitations, graphene has emerged as an alternative
plasmonic material due to its extraordinary electronic and optical
properties~\cite{Graphene1, Graphene2,
grigorenko2012,bao2012,bludov2013,abajoreview}. Indeed, graphene hosts extremely
confined plasmons, facilitating strong light-matter
interactions~\cite{grigorenko2012,bao2012,bludov2013,abajoreview}. In addition,
the plasmon spectrum in doped graphene is highly tunable through electrical or
chemical modification of the charge carrier density. Due to these properties,
graphene is a promising material platform for several photonic applications,
specially in the THz frequency range~\cite{bludov2013}. At the quantum level,
the spatial confinement of surface plasmons in graphene has been shown to
modify the SE rate~\cite{koopens2013,gaudreau2013}. The electromagnetic (EM) field pattern excited by
quantum emitters near a graphene sheet~\cite{hanson2012}
further demonstrates the huge field enhancement due to the excitation of surface
plasmons. A graphene sheet has also been shown to mediate sub- and
super-radiance between two quantum emitters~\cite{huidobro2012}. Recently, the
electrical control of the relaxation pathways and SE rate in graphene has been
observed~\cite{tielrooij2015}. Despite all these advances, the achieved modification 
in the emitter's decay rate  remains modest so far. Most of the proposed schemes
consider emitters whose transition frequencies are in the optical/near infrared range, 
usually far from graphene's intraband transitions. 
As a consequence, the effects of graphene on the SE rate are only relevant when the emitter is
no more than a few dozen nanometers apart.

Here, we propose an alternative mechanism to actively tune the lifetime of a THz quantum emitter near a graphene sheet by exploiting its 
extraordinary magneto-optical properties.  We show that the application of a
magnetic field ${\bf B}$ allows for an unprecedented control of the SE rate for
emitter-graphene distances in the micrometer range. This is in contrast to
previous proposals, in which the 
modification of the SE rate was achieved by electrically or chemically altering graphene's doping level. The fact that we consider a 
low-frequency emitter enables us to probe the effects of intraband transitions in graphene on the decay rate, which have also been unexplored so far. In summary, our key results are (i) a 
striking 99$\%$ reduction of the emitter SE rate compared
to the case where ${\bf B} = {\bf 0}$; (ii) a new distance-scaling
law for the decay rate that corrects the typical $1/d^4$ behavior and is valid
for a broad range of distances and magnetic fields; (iii) a highly non-monotonic
behavior of the SE rate as a function of $|{\bf B}|$, with sharp discontinuities
in the regime of low temperatures; and (iv) the possibility of tailoring the
decay channels into which the photon can be emitted. These findings
can be physically explained in terms of the interplay among the different EM
modes and of electronic intraband transitions between discrete Landau  levels in graphene. 

\section{Methods}

\begin{figure}[h!]
\centering
\includegraphics[scale=0.42]{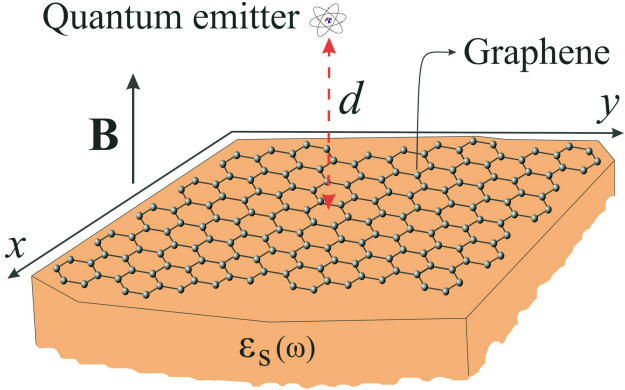}
\vspace{-5pt}
\caption{Quantum emitter at a distance $d$ above a graphene sheet on
the top of a substrate of permittivity
$\varepsilon_s(\omega)$. The whole system is under the
influence of a magnetic field ${\bf B} = B\hat{{\bf z}}$.}
\label{Fig1}
\end{figure}

Let us consider the situation depicted in Fig~\ref{Fig1}.  The half-space $z<0$ is composed of a
non-magnetic, isotropic, and homogeneous material of permittivity
$\varepsilon_s(\omega)$, on top of which ($z=0$) a flat graphene sheet is
placed. The system is under influence of a
uniform static magnetic field ${\bf B} = B\hat{{\bf z}}$. The upper medium $z>0$
is vacuum and an excited quantum emitter is located at a distance $d$ above the interface.   

We consider that the quantum emitter
dynamics is well described by two of its energy eigenstates ($\vert g
\rangle$ and $\vert e \rangle$). 
Within the
electric dipole approximation and weak-coupling regime, one can show
that the SE rates $\Gamma_{\perp}$  and $\Gamma_{\|}$ for transition dipole
moments perpendicular and parallel to the $XY$-plane respectively are \linebreak (Appendix A) \cite{Novotny, Haroche1992, KortKamp15}
\begin{eqnarray}
\label{SEPerp}
\!\!\!\!\dfrac{\Gamma_{\perp}}{\Gamma_0} &=& 1 \! +\! \dfrac{3}{2k_0^3}\
\textrm{Im}\!\left\{i \!\!\int_0^{\infty}\dfrac{k^3 e^{2i{k_z} d}
dk}{{k_z}}r^{\textrm{p,p}}\right\}, \\
\label{SEPar}
\!\!\!\!\dfrac{\Gamma_{\|}}{\Gamma_0} &=& 1 \! +\! \dfrac{3}{4k_0}\
\textrm{Im}\!\left\{i \!\!\int_0^{\infty}\!\dfrac{k e^{2i{k_z} d}
dk}{{k_z}}\left[r^{\textrm{s,s}}-\dfrac{{k_z}^2}{k_0^2}r^{\textrm{p,p}}\right]\!\right\},
\end{eqnarray}
where $\Gamma_0 = |{\bf d}_{\rm eg}|^2 \omega_{0}^3/ (3 \pi \varepsilon_0\hbar c^3)$ is the free space SE rate, $\bf{d}_\textrm{eg}$ is the 
emitter's electric dipole matrix element,  $\omega_0 = k_0c$ is the transition frequency, $k_z = \sqrt{k_0^2-k^2}$, and $r^{\textrm{s,s}}$,
$r^{\textrm{p,p}}$ are the graphene-coated wall polarization preserving reflection coefficients. Although the cross-polarization 
reflection coefficients $r^{\textrm{s,p}}$
and $r^{\textrm{p,s}}$ are non-vanishing in the case of graphene under the influence 
of an uniform static magnetic field, being responsible for Faraday and Kerr rotations, they do not contribute to the 
emitter's lifetime in the present situation (see Appendix A).
The diagonal reflection coefficients are given by  (Appendix B) \cite{WangKong, Kort-Kamp14, KortKamp15}
\bea
r^{\textrm{s,s}} =
-\frac{\Lambda^2+\Delta_{+}^{L}\Delta_{-}^{T}}{\Lambda^2+\Delta_{+}^{L}\Delta_
{ + } ^ {T}} \;\;\; , \;\;\;
r^{\textrm{p,p}} =
\frac{\Lambda^2+\Delta_{-}^{L}\Delta_{+}^{T}}{\Lambda^2+\Delta_{+}^{L}\Delta_{+
} ^ {T} }\, ,
\label{reflectioncoefficients}
\eea
where $\Lambda^2 = Z_0^2k_{z}k_{z}^s \sigma_{H}^2$,
$\Delta_{\pm}^{L} =  k_{z}\varepsilon_{s}/ \varepsilon_{0} \pm k_{z}^{s}+k_{z}k_{z}^s\sigma_{L}/(\omega \varepsilon_0)$,
$\Delta_{\pm}^{T} =  k_{z} \pm k_{z}^{s}+\mu_0 \omega \sigma_{T}$, $Z_0=\sqrt{\mu_0/\varepsilon_0}$, $k_{z}^s =
\sqrt{\mu_0\varepsilon_{s}\omega_0^2 - k^2}$, and $k = |{\bm k}| = |k_x\hat{{\bf x}}+k_y\hat{{\bf y}}|$.  $\sigma_{L}$, $\sigma_{T}$  and
$\sigma_{H}$ are the longitudinal, transverse and Hall
conductivities of graphene, respectivelly, which are in general functions of both frequency and tranverse wavevector ${\bm k}$.  Although the dependence of the material 
properties on wavevector may be relevant in the near-field~\cite{FordWeber}, we have checked that this is not the case for the distances we consider. 
Indeed,  the evanescent waves contribution to the SE process is suppressed by a $e^{-2k d}$ factor,  whereas non-local effects on graphene's 
conductivity become significant for $k \gtrsim \max\left(\sqrt{eB/\hbar},\omega_0/v_F, \tau^{-1}/v_F\right)$ \cite{Nuno-MagnetoPlasmons,Sauber}. 
Here, $v_F \simeq 10^6$ m/s and $\tau$ is a phenomenological relaxation time of electrons in graphene. Therefore, provided $d\gg 
\min\left(\sqrt{\hbar/eB},v_F/\omega_0, v_F\tau\right)$ we can safely set $k=0$ in the 
conductivities, in which 
case $\sigma_{L}=\sigma_{T}$. 

We will study the lifetime of quantum
emitters in the low temperature ($k_BT\ll \mu_c$) and frequency ($\hbar \omega_0 \ll \mu_c$) regimes, where $\mu_c$ is the graphene's chemical 
potential. 
As a result, graphene's conductivities can be approximated by their intraband terms ~\cite{bludov2013,Gusynin1,Gusynin2, Graphene3} 
\begin{eqnarray}
\!\!\!\!\sigma_{L} = \sigma_{T} \simeq \sigma_{L}^{\textrm{intra}}
&\simeq&\dfrac{e^3v_F^2\hbar B
(\omega +
i\tau^{-1})(1+\delta_{0,n_c})}{i\pi\Delta_{\textrm{intra}}[\Delta_{\textrm{intra
}}^2 - \hbar^2(\omega + i\tau^{-1})^2]}\, ,
\label{LongitudinalIntra}\\ \cr
\!\!\!\! \sigma_{H} \simeq \sigma_{H}^{\textrm{intra}} &\simeq&
-\dfrac{e^3v_F^2 B
(1+\delta_{0,n_{\textrm{c}}})}{\pi[\Delta_{\textrm{intra}}^2 - \hbar^2(\omega +
i\tau^{-1})^2]}\, ,
\label{TransversalIntra}
\end{eqnarray}
where $\Delta_{\textrm{intra}}\! =\! M_{n_c+1}\! -\! M_{n_c} $, 
$M_n=\sqrt
{n} M_1$ are the Landau energy levels, $M_1^2 = 2 \hbar e B v_F^2$, and $n_c\!\! =\!
\textrm{int}[\mu_c^2/M_1^2]$ denotes the number of occupied Landau
levels.

\section{Results}

Following previous experimental work on
SE \cite{Hulet-Hilfer-Kleppner-85}, we consider from now on
an emitter with a strong transition at $\omega_0 = 4.2 
\times 10^{12}$ rad/s ($\sim$ 0.7 THz). 
We set  $\tau\! =\! 0.184$ ps
\cite{Ju},  $\mu_c = 115$ meV and, inspired by recent experiments on magneto-optical effects in graphene~\cite{crassee2012}, consider a silicon 
carbide (SiC) substrate.  It is important to clarify that 
$\omega_0$ may be a function of $d$ (the
emitter energy levels can be Lamb-shifted) and of $B$ (the levels
may also be Zeeman-shifted). However, for the purposes of the present
work, both effects may be neglected. A numerical estimate shows that
for the distances considered here, the influence of the Lamb shift   
on the SE rate is unnoticeable,  regardless of the value of $B$. Concerning the Zeeman shift, we have checked that although some
energy levels may be altered in their absolute values, the suppression and enhancement factors of the SE rate due to the application of $B$ are insensitive to this shift.

\begin{figure}
\centering
\includegraphics[scale=0.4]{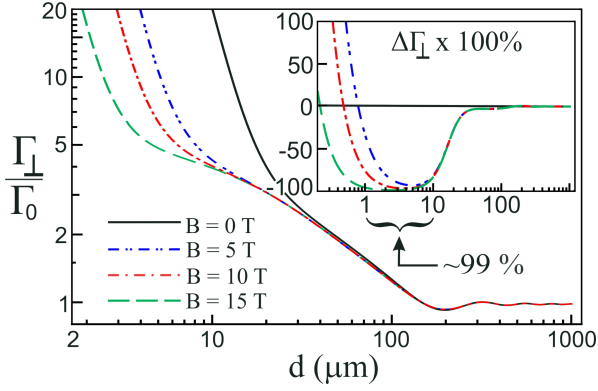}
\vspace{-5pt}
\caption{Normalized decay rate $\Gamma_{\perp}/\Gamma_0$ as a function of distance $d$ between the emitter and the graphene-SiC half-space for
different  magnetic fields. The inset presents the relative SE rate $\Delta\Gamma_{\perp} $ as a function of
$d$ for the same values of $B$.}
\label{Fig2}
\end{figure}

In Fig. \ref{Fig2}  we plot the normalized SE rate $\Gamma_{\perp}/\Gamma_0$ as
a function of the distance $d$ between the emitter and the half-space for
several values of $B$.  For $d \gtrsim 100\ \mu$m  the coupling
between the emitter and the graphene-coated wall is mediated by propagating
modes ($k \leq k_0$) of the vacuum EM field. In this regime of distances
the emitter's lifetime is barely affected by $B$. This
behavior results from the fact that in the far-field the
phase $e^{2ik_z d}$ gives a highly oscillatory integrand in Eq. (\ref{SEPerp}),
except for $k_{z} \sim 0$.  In this case, however,
 $r^{\textrm{s, s}} \sim r^{\textrm{p, p}} \sim -1 +
{\cal{O}}(k_{z}/k_0)$, so that the reflectivity of the graphene-coated
half-space is almost saturated. Hence $B$ hardly
affects the reflection coefficients in this regime.  A transition from the
oscillating pattern at large distances to a sharp growth at small distances
takes place for $d\lesssim 100\ \mu$m. In this regime of distances the emission
is dominated by evanescent modes  ($k > k_0$) of the vacuum EM field.
Interestingly, for $d \lesssim 10\ \mu$m changing $B$
strongly affects the lifetime of the quantum emitter.   A
striking suppression of $99\%$ in the Purcell factor, when compared to the case
where $B = 0$ T, occurs for $1\ \mu\textrm{m}\ \lesssim d \lesssim 10\ \mu$m and  $B \gtrsim
10$ T. Even for smaller values of $B$ the Purcell effect is greatly reduced.
For example, for $d = 3\ \mu$m the influence of the graphene-coated wall
on $\Gamma_{\perp}$ can be reduced by a factor of 10 for $B = 5$ T.
These results are highlighted in the inset of Fig. \ref{Fig2}  where we plot $\Delta\Gamma_{\perp} =
[\Gamma_{\perp}(d,B)-\Gamma_{\perp}(d,0)]/\Gamma_{\perp}(d,0)$ as a function of
$d$ for the same values of $B$. For  even smaller distances an enhancement of the SE 
rate takes place as the magnetic field increases. 
For clarity this effect is not shown in Fig. \ref{Fig2}, although it can 
be noticed in the inset for $d\lesssim 1\ \mu$m.  For instance,
for $B = 5$ T and $d = 0.2\ \mu$m  the SE rate is enhanced by $\sim 500\%$.

It is also interesting to analyze the distance-scaling law of the SE rate for
graphene under an external magnetic field. In the near-field regime, one can
show that (Appendix C)
\begin{equation}
\label{ScalingLaw}
\dfrac{\Gamma_{\perp}}{\Gamma_0}\simeq\dfrac{3\varepsilon_0c^3 
\textrm{Re}[\sigma_{L}]}{\omega_0^4(\varepsilon_s+\varepsilon_0)^2}
 \dfrac{1}{d^4} \
F\!\left(\frac{|\textrm{Im}\sigma_{L}|}{\omega_0(\varepsilon_s+\varepsilon_0)d
} \right) ,
\end{equation}
where $F(x)$ is defined in Appendix C, provided
$\textrm{Re}[\sigma_{L}] \ll \omega_0(\varepsilon_s+\varepsilon_0)d$ and
$\textrm{Im}[\varepsilon_s(\omega_0)]\simeq 0$. The 
validity of this equation is not restricted to the case when a magnetic field is present, rather it is valid whenever correction due to $\text{Im}\ \sigma_L$ are 
non-negligible. Equation (\ref{ScalingLaw}) explains the results in
Fig. \ref{Fig2} for a broad range of distances  ($0.3\ \mu$m $\lesssim d
\lesssim 1.4\ \mu$m) and magnetic fields ($ 5$ T $ \leq B \leq 20$ T)
 with error
$\lesssim 10\%$. The distance scaling-law $\Gamma_\perp \propto
d^{-4} F(d_0/d)$ [where $d_0 =
|\textrm{Im}\sigma_{L}|/\omega_0(\varepsilon_s+\varepsilon_0)$] differs from
the recently observed result $\Gamma_{\perp} \propto d^{-4}$, obtained in the
case ${\bf B} = {\bf 0}$ \cite{gaudreau2013}. This difference arises due 
to (low frequency) intraband transitions and losses in graphene,
 whose signature is coded in the function $F(x)$ appearing in Eq.
(\ref{ScalingLaw}). Indeed, while in the 
high frequency regime ($\omega_0 \gg \tau^{-1}$)  graphene's 
conductivity is approximately a real function, this is not true at the frequency
considered here ($\omega_0 \sim \tau^{-1}$). However, the 
$\Gamma_{\perp} \propto d^{-4}$ can be derived provided $|\textrm{Im}\ \sigma_{L}| \ll \omega_0(\varepsilon_s+\varepsilon_0)d$. Since 
$\textrm{Im}[\sigma_{L}]$ is greatly affected by $B$, the magnetic field could be exploited to tailor the distance ranges where this condition is 
satisfied, allowing for a real time control of the distance-scaling law in the
near-field. Note that the effects of $B$ on the SE are predominantly related to
changes in $\sigma_{L}$.  We have verified that $\sigma_{H}$ can be neglected in
Eq. (\ref{reflectioncoefficients}) for the chosen material parameters. In this
case, the same modifications in the SE rate could be obtained by 
applying a trigonal distortion to graphene, which would generate a strain induced pseudo-magnetic field, 
giving rise to the formation of Landau levels in graphene's electronic spectrum, while keeping $\sigma_H=0$ due to time-reversal invariance. \cite{LevyPseudoMag, GeimPseudo}
\begin{figure}
\centering
\includegraphics[scale=0.41]{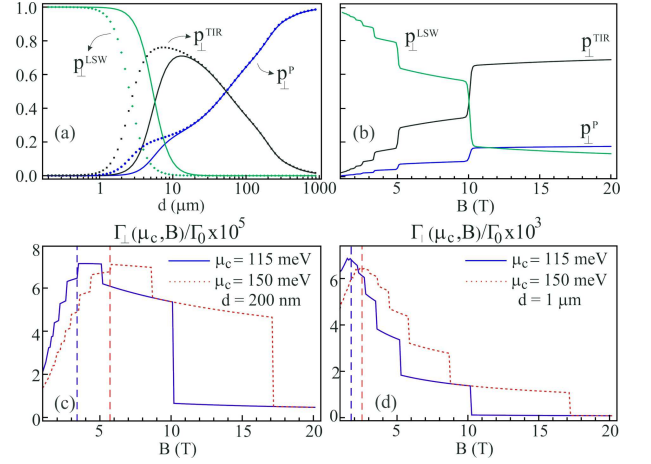}
\vspace{-10pt}
\caption{The decay channel probabilities as a function of {\bf (a)} $d$, and {\bf (b)}
$B$ for $\mu_c = 115$ meV. In (a) solid and dotted curves are for $B = 5$ T and
$B = 15$ T, respectively. In (b) the distance is fixed at $d = 4\ \mu m$.
$\Gamma_{\perp}(d, B)/\Gamma_0$ as a function of $B$ is plotted in
{\bf (c)} $d = 200$ nm,  and {\bf(d)} $d = 1\ \mu$m. The vertical lines show the
position of the peak of $\Gamma_{\perp}(d, B)/\Gamma_0$ obtained via
Eq. (\ref{MaximoCampoMagnetico}).}
\label{Fig3}
\end{figure}

To understand such an influence of $B$ on the SE rate in the
near-field it is necessary to delve a little deeper into the decay process
itself. The spontaneous decay of a source is often associated to the emission of
radiation to the far-field, but that is not necessarily the case. In particular, in the
near-field regime the emitter decays preferentially into non-radiative channels, like surface waves 
characterized  by $k \geq
\sqrt{\varepsilon_s/\varepsilon_0}k_0$ \cite{Note2}. For the transition frequency we are considering, $|\textrm{Re}\ \sigma_{L}| \sim |\textrm{Im}\ \sigma_{L}|$, so 
that surface
magneto-plasmon polaritons~\cite{Nuno-MagnetoPlasmons} are strongly damped, playing essentially no role
in the SE process \cite{footnote}. Nevertheless, the so called lossy
surface waves (LSW) \cite{FordWeber, Lakowicz, Barnes98} are crucial here. These
waves correspond to non-radiative processes and emerge in the case where the emitter's  energy is transferred directly to
the substrate or graphene, generally giving origin to an excitation ({\it e.g.}
electron-hole pair). Such waves are quickly damped,
with their energy being usually converted into heat~\cite{FordWeber, Barnes98}.
In the extreme near-field regime, absorption in the materials governs
the SE process and the LSW ($k \gg k_0$) are usually the main channel into which
the emitter loses its energy. 

In Fig.~\ref{Fig3} we unveil the role played by the different decay
channels in
the emitter's lifetime.  Fig. \ref{Fig3}(a) depicts the decay probabilities
$p^{\textrm{P}}_{\perp}$, $p^{\textrm{TIR}}_{\perp}$, and $p^{\textrm{LSW}}_{\perp}$
of energy emission in a propagating (P), totally internal reflected (TIR), or LSW mode, respectively, as
functions of $d$ for two different values of $B$. These probabilities are given 
by the ratio between the partial decay rates into the aforementioned modes 
and the total SE rate. The partial contribution of propagating, TIR, and LSW modes
to the SE rate can be respectively well approximated by (Appendix D)
\begin{eqnarray}
\label{radiative_allowed}
&\dfrac{\Gamma_{\perp}^{\textrm{P}}}{\Gamma_0} &\simeq 1 \! +\! \dfrac{3}{2}\
\!\!\int_0^{k_0}\dfrac{k^3\ \textrm{Re}[ r^{\textrm{p,p}}e^{2i\sqrt{k_0^2-k^2} d}]
}{k_0^3\sqrt{k_0^2-k^2}}dk\, , \\ \cr
\label{radiative_forbidden}
&\dfrac{\Gamma_{\perp}^{\textrm{TIR}}}{\Gamma_0} &\simeq \dfrac{3}{2} \displaystyle{\int_{k_0}^{k_0\sqrt{\frac{\varepsilon_s}{\varepsilon_0}}}} 
\dfrac{k^3e^{-2\sqrt{k^2 - k_0^2}  z}}{k_0^3\sqrt{k^2 - k_0^2} }  \textrm{Im}[r^{\textrm{p, p}}]  dk \, ,\\ \cr
\label{non_radiative}
&\dfrac{\Gamma_{\perp}^{\textrm{LSW}}}{\Gamma_0} &\simeq  \dfrac{3}{2} \int_{k_0\sqrt{\frac{\varepsilon_s}{\varepsilon_0}}}^{\infty} \dfrac{k^2 
e^{-2k z}  }{k_0^3} \textrm{Im}[r^{\textrm{p, p}}_{\textrm{QS}}] dk \, , 
\end{eqnarray}
where $r^{\textrm{p, p}}_{\textrm{QS}}$ are the reflection coefficients of the graphene-on-substrate system in the quasi-static limit [equivalent to 
take $c \rightarrow \infty$ in Eq. (\ref{reflectioncoefficients})].

We note in Fig. \ref{Fig3} that
changing $B$ can severely affect the possible decay
channels in the $1\ \mu\textrm{m} \lesssim d \lesssim 10\ \mu\textrm{m}$ range,
essentially swapping the role of the LSW and TIR modes as the dominant decay
pathway. Indeed, for $d = 4\ \mu$m we note that $p_{\perp}^{\textrm{LSW}}$ drops
sharply from $75\%$  to $15\%$ when $B$ changes from $5$ T to $15$ T. On the
other hand,  $p_{\perp}^{\textrm{TIR}}$ ($p_{\perp}^{\textrm{P}}$) increases
from $20\%$ ($5\%$) to $67\%$ ($18\%$). This effect is evinced in
Fig.~\ref{Fig3}(b), where we plot the decay probabilities as a function of $B$, for $d = 4\ \mu$m. It is then clear the overall downward
(upward) trend of $p_{\perp}^{\textrm{LSW}}$ ($p_{\perp}^{\textrm{TIR}}$) as $B$
is increased, with a dominance exchange at $B \simeq 10$ T.   

Figures \ref{Fig3}(c) and \ref{Fig3}(d) show $\Gamma_{\perp}(d,
B)/\Gamma_0$ as a function of $B$ for $d = 200$ nm and $d = 1\
\mu$m, respectively, and two distinct values of $\mu_c$. The SE
rate presents sharp discontinuities, which are directly linked to the discrete
character of the 
Landau levels brought about by the application of ${\bf B}$. These
discontinuities occur whenever a given Landau level energy crosses $\mu_c$
\cite{Kort-Kamp14,Gusynin1,Gusynin2, Graphene3}. Moreover, there exists a critical magnetic
field $B_c = \mu_c^2/(2\hbar e v_F^2)$ above which the discontinuities are no longer present. This is due to the fact that for $B > B_c$ all positive
Landau levels are above $\mu_c$, so no more crossings can occur. Note that the curves merge in the final plateau, regardless of the
value of $\mu_c$.  For $B  > B_c$ we have $\Delta_{\textrm{intra}} =
M_1$, that does not depend on $\mu_c$. Hence,  provided $k_BT \ll \mu_c$
both $\sigma_{L}$ and $\sigma_{H}$ are approximately independent of $\mu_c$
for $B  > B_c$.
\begin{figure}
\centering
\includegraphics[scale=0.35]{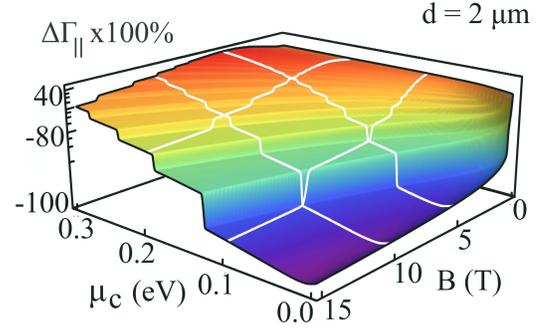}
\caption{3D plot of the relative spontaneous emission $\Delta\Gamma_{{\small
|\!|}}(\mu_c,B)$ as a function of both $\mu_c$ and $B$ for $d = 2\, \mu$m.}
\label{Fig4}
\end{figure}
As a function of $B$, the SE rate presents a
maximum whose position depends on both $\mu_c$ and $d$.
This behavior can be understood recalling 
that for short distances the SE rate is \cite{KortKamp15, gaudreau2013}
\begin{equation}
\dfrac{\Gamma_{\perp}}{\Gamma_0} \simeq \dfrac{3}{2 k_0^3} \int_{0}^{\infty} d
k \rho(k)\,\textrm{Im}\left[r^{\textrm{p, p}}(k,\omega_0,B) \right]\, ,
 \label{SER_Perp_Approx}
\end{equation}
where $\rho(k) = k^2 e^{-2k d}$ has a maximum at
$k^{\textrm{max}}_{1} = 1/d$.  In the large $k$ limit $\textrm{Im}[r^{\textrm{p, p}}]$ presents a
peak at 
$ {k}^{\textrm{max}}_{2}\ \simeq \varepsilon_0 \omega_0
[\varepsilon_{s}/\varepsilon_0+1]/|\sigma_{L}|$. 
Since $|\sigma_{L}(\omega_0, B)|$ decreases with $B$ (for $B
> 1$ T in our case) we note that $k^{\textrm{max}}_{2}$ moves to high
values of $k$ as $B$ increases. Therefore, for a fixed
emitter-graphene separation, the overlap between $\rho(k)$ and 
$\textrm{Im}[r^{\textrm{p, p}}]$ grows with $B$ until ${k_{2}^{\textrm{max}}} \sim
{k_{1}^{\textrm{max}}}$. After that, this overlap diminishes and so does
$\Gamma_{\perp}$, which explains the behavior of the SE rate in Fig.
\ref{Fig3}. The value of the magnetic field
$B_m$ that maximizes $\Gamma_{\perp}$ can be estimated by setting $k^{\textrm{max}}_{1} = k^{\textrm{max}}_{2}$. \linebreak This leads to 
\begin{equation}
|\sigma_{L}(\omega_0, \mu_c, B_m)| \simeq \varepsilon_0 \omega_0 d
\left[\varepsilon_{s}(\omega_0)/\varepsilon_0+1 \right]\, .
\label{MaximoCampoMagnetico}
\end{equation}
The accuracy of this equation is clearly seen in Figs. \ref{Fig3}(c)-(d) where we show $B_m$ calculated through Eq.
(\ref{MaximoCampoMagnetico}) for $\mu_c = 115$ meV and $\mu_c = 150$ meV.

Similar results hold for $\Gamma_{\|}$ in the near-field regime. Indeed,
for $d\ll2\pi/k_0$ the contribution of $r^{\textrm{s, s}}$ to $\Gamma_{\|}$ is negligible and the approximation $k_z \simeq ik$ is valid. 
Hence,  apart from a factor $1/2$,  $\Gamma_{\|}$ can also be written as in Eq. (\ref{SER_Perp_Approx}) (see Appendix C) \cite{gaudreau2013}. 
In Fig. \ref{Fig4} we plot 
$\Delta\Gamma_{{\small \|}}(\mu_c,B) = [\Gamma_{{\small \|}}(\mu_c,
B)-\Gamma_{{\small \|}}(\mu_c, 0)]/\Gamma_{{\small \|}}(\mu_c, 0)$ as a
function of both $\mu_c$ and $B$ for $d = 2\ \mu$m. In this case, the reduction
in the Purcell factor in the $\mu$m range can be as high as $98\%$, when compared to the case ${\bf B} = {\bf 0}$. 
Figure \ref{Fig4} corroborates our conclusions that an astounding control on the radiative properties
of quantum emitters can be achieved via magneto-optical properties in
graphene.  Moreover, Fig. \ref{Fig4} reveals that the SE rate can be
modified by keeping $B$ constant while changing $\mu_c$, which could be implemented by applying a gate voltage on graphene \cite{Graphene1, Graphene2, bludov2013, grigorenko2012,
bao2012}. 

\section{Conclusion}

In conclusion, we have shown that the application of a magnetic field allows for a great control over the Purcell effect and  
decay pathways of quantum emitters near graphene.
Altogether, our findings demonstrate the viability of actively dictating optical energy transfer processes with magnetic 
fields or strain. By demonstrating that these results are within the reach of state-of-the-art experiments on quantum emission in 
the THz range, we expect that they may find further applications in quantum photonics, and may even serve to probe other light-matter phenomena.

\section*{Acknowledgements}

We would like to thank D. A. R. Dalvit, E. C. Marino, F. Guinea, H. Ulbricht, and L. Sapienza  for valuable comments.
W.J.M.K-K., G.B., F.S.S.R., and C.F. acknowledge CNPq, CAPES, and FAPERJ for
financial support. W.J.M.K.-K. acknowledges  financial support
from the LANL LDRD program. B.A. acknowledges financial support from Funda\c{c}\~{a}o para a Ci\^encia e a Tecnologia, Portugal, through Grant No. 
SFRH/BD/78987/2011. F.A.P. thanks the Optoelectronics Research Centre and Centre for Photonic Metamaterials, University of Southampton, for the 
hospitality, and CAPES for funding his visit (Grant No. BEX 1497/14-6). F.A.P. also acknowledges CNPq (Grant No. 303286/2013-0) for financial 
support. N.M.R.P. acknowledges financial support from the Graphene Flagship Project (Contract No. CNECT-ICT-604391).

\section*{Appendix A: Spontaneous emission near an anisotropic interface}
\label{AppendixA}

The spontaneous emission decay rate of a two level emitter can be written in terms of the EM dyadic Green's function as
\begin{equation}
\label{TaxaEmissaoEspontanea3}
\Gamma = \dfrac{2\omega_0^2}{\varepsilon_0 \hbar c^2} \textrm{Im} \left[ \bf{d}_\textrm{ge}^{*} 
\cdot \mathbb{G}({\bf r}_0, {\bf r}_0; \omega_0) \cdot \bf{d}_\textrm{ge} \right]\, ,
\end{equation}
where ${\bf r}_0=(0,0,d)$ is the position of the quantum emitter and $\mathbb{G}({\bf r}, {\bf r}'; \omega)$ is the  EM dyadic Green's function, 
which 
allows one to write the electric field at position ${\bf 
r}$ that is generated by a point dipole at position ${\bf r}'$, oscillating with frequency $\omega$ as
\begin{equation}
\label{Green_dipole}
{\bf E}({\bf r}; \omega) = \mu \omega^{2} \mathbb{G}(\bf{r},\bf{r}'; \omega) \cdot \bf{d}(\omega),
\end{equation}
where $\mu$ is the permeability of the medium where the dipole is embedded. In vacuum the dyadic 
Green's function satisfies the 
inhomogeneous Helmholtz equation
\begin{equation}
\label{HelmholtzGreenFunction}
\nabla \times \nabla \times \mathbb{G}({\bf r}, {\bf r}'; \omega) - \dfrac{\omega^2}{c^2} \mathbb{G}({\bf r}, {\bf r}'; \omega) = \mathbb{I} 
\delta({\bf r} - {\bf r}')\, ,
\end{equation}
with $\mathbb{I}$ being the unit dyad. 

We are interested in the case where the emitter is located at a distance $d$ above a semi-infinite homogeneous medium with 
flat surface at $z = 0$, where the graphene layer lies. 
Since the dyadic Green's function obeys the same boundary conditions as the electric field, we can write it for $z,~z'>0$ as
\begin{eqnarray}
\label{GreenFunctionSplit}
\mathbb{G}({\bf r}, {\bf r'}; \omega_0)  = \mathbb{G}^{\textrm{(0)}}({\bf r}, {\bf r'}; \omega_0) +\mathbb{G}^{\textrm{(r)}}({\bf r}, {\bf r'}; 
\omega_0)\, ,
\end{eqnarray}
where $\mathbb{G}^{\textrm{(0)}}({\bf r}, {\bf r'}; \omega_0)$  is the free space Green's function and $\mathbb{G}^{\textrm{(r)}}({\bf r}, 
{\bf r'}; \omega_0)$ is the reflected one. For $z'>0$ and $z<0$, the dyadic Green's function can be written as a transmitted Green's function, 
$\mathbb{G}^{\textrm{(t)}}({\bf r}, {\bf r'}; \omega_0)$. Each of these Green's functions can be conveniently expressed in terms of its spatial 
2D-Fourier transform $\tilde{\mathbb{G}}^{\textrm{(0/r/t)}}({\bm k},z,z'; \omega)$ as \cite{Novotny}
\begin{eqnarray}
\label{GreenFunction}
\!\!\!\!\!\!\!\!\mathbb{G}^{\textrm{(0/r/t)}}({\bf r}, {\bf r'}; \omega)\! =\!  \int\! \dfrac{d^2 \boldsymbol{k}}{(2\pi)^2} e^{i{\boldsymbol{k}}\cdot ({\bm x} - {\bm 
x}')} \tilde{\mathbb{G}}^{\textrm{(0/r/t)}}({\bm k},z,z'; \omega),
\end{eqnarray}
where $\bm{k}=k_x\hat{{\bf x}}+k_y\hat{{\bf y}}$ and $\bm{x}=x\hat{{\bf x}}+y\hat{{\bf y}}$. The free space Green's function is given by
\begin{eqnarray}
\label{FreeSpaceGreenFunction}
\!\!\!\!\!\!\!\!\!\!\tilde{\mathbb{G}}^{\textrm{(0)}}({\bm k},z,z'; \omega)\! =\! \dfrac{i}{2 k_z}e^{i k_z |z-z'|} \left( {\bm \epsilon}_{\rm p}^{\pm}\otimes {\bm \epsilon}_{\rm p}^{\pm} +  {\bm \epsilon}_{\rm 
s}^{\pm}\otimes {\bm \epsilon}_{\rm s}^{\pm} \right)\, , 
\end{eqnarray}
with $k_z$ defined as
\begin{equation}
\label{def_k_z}
{k_z} =\begin{cases}
\sqrt{k_{0}^{2}-k^{2}} & ,\, k<k_{0}\\
i\sqrt{k^{2}-k_{0}^{2}} & ,\, k>k_{0}
\end{cases},
\end{equation}
and we have introduced the polarization vectors for $s$- and $p$-polarized waves (the $+$ and $-$ signs correspond to $z>z'$ and $z<z'$, respectively)
\begin{align}
\label{polarization_vectors}
\bm{\epsilon}_{\rm s}^{\pm} = \frac{k_y \hat{\bf x}-k_x \hat{\bf y}}{k}, && \bm{\epsilon}_{\rm p}^{\pm} = \frac{k}{k_0} \hat{\bf z} \mp 
\frac{k_z}{k_0}\frac{k_x \hat{\bf x} + k_y \hat{\bf y}}{k}.
\end{align}
Note that these vectors are orthogonal, but they are normalized only for propagating modes ($k < k_0$). 

The reflected Green's function can be written as 
\begin{eqnarray}
\label{ReflectedGreenFunction}
\!\!\!\!\tilde{\mathbb{G}}^{\textrm{(r)}}({\bm k},z,z'; \omega)  = \dfrac{i}{2 k_z}e^{i k_z (z+z')} \sum_{{\rm i, j} = {\rm s, p}} r^{\rm i, j} \bm{\epsilon}_{{\rm i}}^{+} \otimes 
\bm{\epsilon}_{{\rm j}}^{-}\, , 
\end{eqnarray}
where $r^{\rm{i},\rm{j}}$ are the reflection coefficients for an incoming $\rm{j}$-polarized wave that is reflected as an 
$\rm{i}$-polarized wave. Similarly, the transmitted Green's function is given by
\begin{eqnarray}
\label{TransmittedGreenFunction}
\!\!\!\!\!\! \tilde{\mathbb{G}}^{\textrm{(t)}}({\bm k},z,z'; \omega) = \dfrac{i}{2 k_z}e^{-i k_z^s z} e^{i k_z z'}\!\! \sum_{{\rm i, j} = {\rm s, p}} t^{\rm i, j} \bm{\epsilon}_{{\rm i}, t}^{-} \otimes 
\bm{\epsilon}_{{\rm j}}^{-}\, , 
\end{eqnarray}
where $t^{\rm{i},\rm{j}}$ are the transmission coefficients (incoming ${\rm j}$-polarized wave, transmitted ${\rm i}$-polarized wave) and 
$\bm{\epsilon}_{{\rm i}, t}^{\pm}$ are the polarization vectors in the substrate, given by Eq.~\eqref{polarization_vectors} 
after replacing $k_0$ by $k_0 \sqrt{\varepsilon_s/\varepsilon_0}$ and $k_z$ by $k_z^s$. The reflection and transmission 
coefficients are obtained by imposing the usual boundary conditions on the EM field at $z=0$ and by modelling graphene as a two-dimensional current 
distribution \linebreak (see Appendix B). 

The evaluation of the SE rate requires the evaluation of the dyadic Green's function at the coincidence ${\bf r}' = {\bf r} = {\bf r}_0$. In this 
case the integration over the momentum angular variable in Eq.~\eqref{GreenFunction} can be easily performed. The only nonzero components of  
$\mathbb{G}^{\textrm{(0)}}({\bf r}_0, {\bf r}_0; \omega_0)$ are the diagonal ones. The contribution of
$\mathbb{G}^{\textrm{(r)}}({\bf r}_0, {\bf r}_0; \omega_0)$ to the SE rate presents
polarization preserving terms (which involve ${\bm \epsilon}_{\rm p}^{+} \otimes {\bm \epsilon}_{\rm p}^{-}$  and ${\bm \epsilon}_{\rm 
s}^{+} \otimes {\bm \epsilon}_{\rm s}^{-}$)  and cross-polarization terms (which involve ${\bm \epsilon}_{\rm p}^{+} \otimes {\bm \epsilon}_{\rm 
s}^{-}$ and ${\bm \epsilon}_{\rm s}^{+} \otimes {\bm \epsilon}_{\rm p}^{-}$). After performing the angular integration, the polarization preserving 
terms only select the diagonal terms of $\bf{d}_{\rm{ge}}^{*}\otimes \bf{d}_{\rm{ge}}$. The cross polarization terms select the 
$d_{\rm{ge},x}^{*}d_{\rm{ge},y}-d_{\rm{ge},y}^{*}d_{\rm{ge},x}$ components of the dipole matrix elements. As the transition dipole matrix elements of a two-level system
can be made real by a proper choice of the relative phase between $|{\rm g}\rangle$ and $|{\rm e}\rangle$, the cross polarization terms do not contribute to the SE rate. Therefore,  only the reflection 
coefficients $r^{\rm{p,p}}$ and $r^{\rm{s,s}}$ give a non-vanishing contribution to the SE process even though cross polarization coefficients $r^{\rm{s,p}}$ and $r^{\rm{p,s}}$ are non zero. 
Similar conclusions hold even in the case of a semi-infitite anisotropic substrate. By plugging 
Eqs. (\ref{GreenFunctionSplit})-(\ref{ReflectedGreenFunction}) into (\ref{TaxaEmissaoEspontanea3}) it is straightforward to obtain Eqs. (\ref{SEPerp}) and 
(\ref{SEPar}).

\section*{Appendix B: Fresnel's coefficients for an interface coated with a 2D conductive film in the presence of an applied magnetic field}
\label{AppendixB}

Let us consider that an incoming arbitrarily polarized EM wave propagating in a dielectric medium with permittivity $\varepsilon_1$ and permeability 
$\mu_1$, impinges on the flat interface with a second homogeneous medium, with permittivity $\varepsilon_2$ and permeability 
$\mu_2$,  occupying the half-space $z\leq0$ coated by a 2D conductive film.
For an impinging electromagnetic wave with frequency $\omega$ and in-plane wavevector $\bm k$,  the electric and magnetic fields 
can be expressed as
\begin{align}
\label{EletricoIncidente}
{\bf E}_I & = \left[E_I^{\textrm{s}} \mbox{{\mathversion{bold}${\epsilon}$}}_{\textrm{s},1}^{+}  + E_I^{\textrm{p}} 
\mbox{{\mathversion{bold}${\epsilon}$}}_{\textrm{p},1}^{+}  \right]e^{-i k_{z,1} z} e^{i({\bm k} \cdot {\bm x} - \omega t)}\, , \\
\label{MagneticoIncidente}
{\bf H}_I & = \frac{1}{Z_1} \left[ E_I^{\textrm{p}}  \mbox{{\mathversion{bold}${\epsilon}$}}_{\textrm{s},1}^{+}  - 
E_I^{\textrm{s}} \mbox{{\mathversion{bold}${\epsilon}$}}_{\textrm{p},1}^{+}  \right] e^{-i k_{z,1} z} e^{i({\bm k} \cdot {\bm x} - \omega t)} \, ,
\end{align}
where $E_I^{\textrm{s}}, \, E_I^{\textrm{p}}$ are the transverse electric and transverse magnetic incoming amplitudes, respectively. $k_{z,n}$ and 
$\bm{\epsilon}_{s/p,n}^{\pm}$ are given by Eqs.~\eqref{def_k_z} and \eqref{polarization_vectors} replacing $k_0$ by $k_n=\omega \sqrt{\varepsilon_n 
\mu_n}$ with $n=1,2$. $Z_n=\sqrt{\mu_n/\varepsilon_n}$ is the impedance of medium $n$. Similarly, the reflected and transmitted fields are written as
\begin{align}
\label{EletricoRefletido}
{\bf E}_R & = \left[E_R^{\textrm{s}} \mbox{{\mathversion{bold}${\epsilon}$}}_{\textrm{s},1}^{-}  + E_R^{\textrm{p}} 
\mbox{{\mathversion{bold}${\epsilon}$}}_{\textrm{p},1}^{-}  \right] e^{i k_{z,1} z} e^{i({\bm k} \cdot {\bm x} - \omega t)}\, , \\
\label{MagneticoRefletido}
{\bf H}_R & = \frac{1}{Z_1} \left[ E_R^{\textrm{p}}  \mbox{{\mathversion{bold}${\epsilon}$}}_{\textrm{s},1}^{-}  - 
E_R^{\textrm{s}} \mbox{{\mathversion{bold}${\epsilon}$}}_{\textrm{p},1}^{-}  \right] e^{i k_{z,1} z} e^{i({\bm k} \cdot {\bm x} - \omega t)} \, ,
\end{align}
and
\begin{align}
\label{EletricoTransmitido}
{\bf E}_T & = \left[E_T^{\textrm{s}} \mbox{{\mathversion{bold}${\epsilon}$}}_{\textrm{s},2}^{+}  + E_T^{\textrm{p}} 
\mbox{{\mathversion{bold}${\epsilon}$}}_{\textrm{p},2}^{+}  \right] e^{-i k_{z,2} z} e^{i({\bm k} \cdot {\bm x} - \omega t)}\, , \\
\label{MagneticoTransmitido}
{\bf H}_T & = \frac{1}{Z_2} \left[ E_T^{\textrm{p}}  \mbox{{\mathversion{bold}${\epsilon}$}}_{\textrm{s},2}^{+}  - 
E_T^{\textrm{s}} \mbox{{\mathversion{bold}${\epsilon}$}}_{\textrm{p},2}^{+}  \right] e^{-i k_{z,2} z} e^{i({\bm k} \cdot {\bm x} - \omega t)} \, .
\end{align}

We should determine the reflected $E_R^{\textrm{s (p)}}$ and transmitted $E_T^{\textrm{s (p)}}$ amplitudes in order to calculate the reflection and 
transmission coefficients
\begin{equation}
r^{\textrm{i, j}} = \dfrac{E_R^{\textrm{i}}}{E_I^{\textrm{j}}}\ \ \ \textrm{and} \ \ \  t^{\textrm{i, j}} = \dfrac{E_T^{\textrm{i}}}{E_I^{\textrm{j}}}\, , \ \ \ \ \textrm{(i, j)} = \textrm{(s, p)}\, .
\end{equation}

The reflected and transmitted amplitudes are obtained by solving Maxwell's equations and imposing the appropriate boundary conditions on the interface 
at $z = 0$.  Taking into account the presence of a 2D conductive film at the $z=0$, the boundary conditions that must be satisfied by the EM field 
are
\begin{align}
\label{CC1}
{\bf \hat{z}} \times \left[{\bf E}_T - {\bf E}_R - {\bf E}_I \right] & = {\bf 0}\, , \\
\label{CC2}
{\bf \hat{z}} \times \left[{\bf H}_T - {\bf H}_R - {\bf H}_I \right] & = {\bf J}_{\rm 2D} =  \mbox{{\mathversion{bold}${\sigma}$}}\cdot{{\bf E}_T}\, 
,
\end{align}
where ${\bf J}_{\rm 2D}$ is a 2D current density that is induced on the conductive field, and $\bm{\sigma}$ is 
the 2D conductivity tensor of the film \cite{note3}. In the most general case (a 2D homogeneous anisotropic material in the presence of a magnetic 
field) the conductivity tensor can be written as  
\begin{align}
\bm{\sigma} & = \sigma_{L} \hat{\bf e}_\parallel \otimes \hat{\bf e}_\parallel + \sigma_{T} \hat{\bf 
e}_\perp \otimes \hat{\bf e}_\perp \\ \nonumber
            & + \sigma_{H}(\hat{\bf e}_\perp \otimes \hat{\bf e}_\parallel - \hat{\bf e}_\parallel \otimes \hat{\bf e}_\perp) \\ \nonumber
            & + \sigma^{\rm sym}_{xy}(\hat{\bf e}_\perp \otimes \hat{\bf e}_\parallel + \hat{\bf e}_\parallel \otimes \hat{\bf e}_\perp),
\end{align}
where $\hat{\bf e}_\parallel = (k_x \hat{{\bf x}} + k_y\hat{{\bf y}})/ \left|\bm{k} \right|$ and $\hat{\bf e}_\perp = (k_y \hat{{\bf x}} - k_x\hat{{\bf y}})/ \left|\bm{k} \right|$.  
$\sigma_{L}$ ($\sigma_T$) is the longitudinal (transverse) conductivity, $\sigma_H$ is the Hall conductivity and $\sigma_{xy}^{\rm sym}$ is only nonzero in 
anisotropic materials such as black phosphorus \cite{Low2014}. In  case of graphene we have $ \sigma^{\rm sym}_{xy} = 0$, but in order to keep the 
discussion as general as possible and due to the rising interest in black phosphorus we will allow for a finite  $\sigma^{\rm sym}_{xy}$.
Using Eqs. 
(\ref{EletricoIncidente})-(\ref{MagneticoTransmitido}) into Eq. (\ref{CC1}) and Eq. (\ref{CC2}) one can demonstrate that the reflected and transmitted 
amplitudes satisfy \linebreak the following equations
\begin{eqnarray}  
E_{I}^{{\rm s}}+E_{R}^{{\rm s}} &=&  E_{T}^{{\rm s}}  ,\\
\dfrac{k_{z,1}}{k_{1}}\left(E_{I}^{{\rm p}}-E_{R}^{{\rm p}}\right)   &=& \dfrac{k_{z,2}}{k_{2}}E_{T}^{{\rm p}},\\
 \dfrac{1}{Z_{1}}\dfrac{k_{z,1}}{k_{1}}\left(E_{I}^{{\rm s}}-E_{R}^{{\rm s}}\right) &=& 
\left(\sigma_{T}+\dfrac{1}{Z_{2}}\dfrac{k_{z,2}}{k_{2}}\right)E_{T}^{{\rm s}} \cr
&+& \left( \sigma_{xy}^{{\rm sym}}+\sigma_{H} \right)\dfrac{k_{z,2}}{k_{2}}E_{T}^{{\rm p}}, \\
\dfrac{1}{Z_{1}}\left(E_{I}^{{\rm p}}+E_{R}^{{\rm p}}\right)   &=&  \left(\sigma_{L}\dfrac{k_{z,2}}{k_{2}}+\frac{1}{Z_{2}}\right)E_{T}^{{\rm  p}}\cr 
&+&\left(\sigma_{xy}^{{\rm sym}}-\sigma_{H}\right)E_{T}^{{\rm s}}.
\end{eqnarray}

Considering separately the cases of s and p incident polarization one can decouple previous equations and show that Fresnel's coefficients in  
the presence of an external magnetic field are given as
\begin{align}
\!\!\!\!\!r^{{\rm p,p}} & = \dfrac{\Delta_{+}^{T}\Delta_{-}^{L}+\Lambda^{2}}{\Delta_{+}^{T}\Delta_{+}^{L}+\Lambda^{2}}, \;\;\;\;\;\;\;\;\;\; r^{{\rm 
s,s}}=-\dfrac{\Delta_{-}^{T}\Delta_{+}^{L}+\Lambda^{2}}{\Delta_{+}^{T}\Delta_{+}^{L}+\Lambda^{2}},\\
\!\!\!\!\!t^{{\rm p,p}} & = \dfrac{Z_{2}\varepsilon_{2}}{Z_{1}\varepsilon_{0}}\frac{2k_{z,1}\Delta_{+}^{T}}{\Delta_{+}^{T}\Delta_{+}^{L}+\Lambda^{2}},\;\; 
t^{{\rm s,s}}=\dfrac{\mu_{2}}{\mu_{0}}\frac{2k_{z,1}\Delta_{+}^{L}}{\Delta_{+}^{T}\Delta_{+}^{L}+\Lambda^{2}},\\
\!\!\!\!\!r^{{\rm s,p}} & = t^{{\rm s,p}}=-2\dfrac{Z_{0}^{2}}{Z_{1}}\frac{\mu_{1}\mu_{2}}{\mu_{0}^{2}}\dfrac{k_{z,1}k_{z,2}\left(\sigma_{xy}^{{\rm 
sym}}+\sigma_{H}\right)}{\Delta_{+}^{T}\Delta_{+}^{L}+\Lambda^{2}},\\
\!\!\!\!r^{{\rm p,s}} & \!=\! -\dfrac{k_{1}k_{z,2}}{k_{2}k_{z,1}}\ t^{{\rm 
p,s}}\!=\!2\dfrac{Z_{0}^{2}}{Z_{1}}\dfrac{\mu_{1}\mu_{2}}{\mu_{0}^{2}}\dfrac{k_{z,1}k_{z,2}\left(\sigma_{xy}^{{\rm 
sym}}\!-\!\sigma_{H}\right)}{\Delta_{+}^{T}\Delta_{+}^{L}+\Lambda^{2}},
\end{align}
with 
\begin{align}
\Delta_{\pm}^{L} & =  \left(k_{z,1}\varepsilon_{2}\pm k_{z,2}\varepsilon_{1}+k_{z,1}k_{z,2}\sigma_{L}/\omega\right)/\varepsilon_{0}, \\
\Delta_{\pm}^{T} & =  \left(k_{z,2}\mu_{1}\pm k_{z,1}\mu_{2}+\omega\mu_{1}\mu_{2}\sigma_{T}\right)/\mu_{0}, \\
\Lambda^{2} & =  Z_{0}^{2}\mu_{1}\mu_{2}k_{z,1}k_{z,2}\left[\sigma_{H}^{2}-(\sigma_{xy}^{\rm sym})^{2}\right]/\mu_{0}^{2}.
\end{align}
For graphene $\sigma_{xy}^{\rm sym}=0$ and, in the case where medium 1 is vacuum ($\varepsilon_1 = \varepsilon_0$, $\mu_1 = \mu_0$) and  
medium 2 is non-magnetic ($\mu_2 =\mu_0$), the reflection coefficients reduce to the ones given in the main text,  Eq.~\eqref{reflectioncoefficients}.
\section*{Appendix C: Distance-scaling law in the near field for terahertz emitters}
\label{AppendixC}
In the near field, the main contribution for the SE rate in Eqs.~\eqref{SEPerp} and \eqref{SEPar} comes from large in-plane wavevectors $k \gg k_0$. In this case the quasi-static approximation holds \linebreak $(c \rightarrow \infty)$ and  $k_z$ and $k_z^{s}$ can be well approximated by $i k$. Besides, the Hall conductivity gives a negligible contribution to the quasi-static reflection coefficients so that we can set $\sigma_H\simeq0$. Within these approximations the dominant terms in $\Gamma_{\perp}$ and $\Gamma_{\|}$ originate from the polarization preserving transverse magnetic reflection coefficient and can be cast as  
\begin{eqnarray}
\label{SEPerpQS}
 \dfrac{\Gamma_\perp}{\Gamma_{0}}     & \simeq& \dfrac{3}{2}\int_{0}^{+\infty}\!\!\! dk \dfrac{k^2}{k_0^{3}} e^{-2 k d}\ {\rm Im} [r_{\rm QS}^{\rm{p,p}}]\, , \\ \cr 
 \label{SEParQS}
 \dfrac{\Gamma_\parallel}{\Gamma_{0}} & \simeq&  \dfrac{3}{4}\int_{0}^{+\infty}\!\!\! dk \dfrac{k^2}{k_0^{3}} e^{-2 k d}\ {\rm Im} [r_{\rm QS}^{\rm{p,p}}]\, , 
\end{eqnarray}
where
\begin{equation}
r_{\rm{QS}}^{\rm{p,p}} = \frac{ i \left(\varepsilon_s - \varepsilon_0\right) \omega_0 - k \sigma_L}{ i \left( 
\varepsilon_0+\varepsilon_s \right) \omega_0 - k \sigma_L}. 
\end{equation}

In a regime where $|{\rm Re}[\sigma_L]| \ll (\varepsilon_0+\varepsilon_s)\omega_0 d$ the imaginary part of $r_{\rm{QS}}^{\rm{p,p}} $ can be  approximated by
\begin{equation}
\label{ReflectionQuasiStatic}
 {\rm Im} [r_{\rm QS}^{\rm{p,p}}] \simeq \frac{ 2 \varepsilon_0 \omega_0 k {\rm Re}[\sigma_L]}{\left\{(\varepsilon_s + \varepsilon_0)\omega_0 - 
k {\rm Im} [\sigma_L] \right\}^{2}}.
\end{equation}

Substituting Eq. (\ref{ReflectionQuasiStatic}) into Eqs. (\ref{SEPerpQS}) and (\ref{SEParQS}) one can show that the Purcell factor in the near-field regime is given by
\begin{align}
 \frac{\Gamma_\perp}{\Gamma_{0}}  & \simeq  \dfrac{3 \varepsilon_0 c^3 {\rm 
Re}[\sigma_L]}{(\varepsilon_0+\varepsilon_s)^{2} \omega_0^{4}}\frac{1}{d^4}\ F\! \left(\frac{|{\rm Im}[\sigma_L]|}{(\varepsilon_0+\varepsilon_s)\omega_0 
d} \right),\\
 \frac{\Gamma_\parallel}{\Gamma_{0}}  & \simeq \frac{3 \varepsilon_0 c^3 {\rm 
Re}[\sigma_L]}{2(\varepsilon_0+\varepsilon_s)^{2} \omega_0^{4}}\frac{1}{d^4}\ F\! \left(\frac{|{\rm Im}[\sigma_L]|}{(\varepsilon_0+\varepsilon_s)\omega_0 
d} \right), 
\end{align}
with the function $F(x)$ defined as
\begin{equation}
 F(x)=\int_{0}^{+\infty} \!\!\!\! dy \frac{y^{3} e^{-2y}}{(1+y x)^2}.
\end{equation}
\section*{Appendix D: Decay channel probabilities}
\label{AppendixD}

In order to determine the different decay pathways probabilities one must study how the total power emitted is distributed into the different channels. Two processes can be distinguished: $(i)$ radiative decay, which involves the 
emission of a photon that can be detected by a far away detector; $(ii)$ non-radiative decay, where the emitted power does 
not reach the far-field, but is instead absorbed by graphene or substrate and creates a material excitation.  In order to compute each channel contribution to the decay process 
we use the fact  that the total SE rate given by Eq. (\ref{TaxaEmissaoEspontanea3}) corresponds also to the power emitted by a classical 
oscillating dipole, ${\bf d}(t)= {\bf d} e^{-i \omega_0 t} + {\bf d}^{*} e^{i \omega_0 t}$. The classical power emitted by such dipole is related to 
the SE rate of a two-level quantum emitter through
$P=\hbar \omega_0\Gamma$ \cite{Novotny, Haroche1992}, provided we choose for ${\bf d}$ the transition dipole moment of the quantum emitter. The probability of decaying into a radiative or non-radiative channel can be obtained by computing the fraction of the power that is emitted by the classical dipole to the far field and the one that is dissipated into the materials, respectively.

The average power emitted by the classical dipole that reaches the far field (radiative processes) can be expressed as
\begin{equation}
\label{AveragePowerDef}
 P_{\rm rad} = \lim_{r\rightarrow \infty}\int_{0}^{2\pi} d\phi \int_{0}^{\pi} d\theta \sin(\theta)\  r^{2} \hat{\bf r}\cdot \langle {\bf 
S}({\bf r})  \rangle , 
\end{equation}
where the Poynting vector in the far field is given by
\begin{equation}
\label{Poyntingvector}
{\bf S}({\bf r},t)=\frac{1}{Z} {\bf E}({\bf r},t)\cdot{\bf E}({\bf r},t) \hat{\bf r},
\end{equation}
and $\langle ... \rangle$ denotes time average over one oscillation period. Here,  $Z=\sqrt{\mu/\varepsilon}$ is the impedance of the medium. Using 
Eqs. ~\eqref{Green_dipole} and (\ref{Poyntingvector}), the time averaged Poynting vector \linebreak can be cast as
\begin{equation}
\label{FF_Poynting}
\langle {{\bf S}}({\bf r},t) \rangle= \frac{2\mu_{0}^2 \omega_0^{4}}{Z} \left| \mathbb{G}({\bf r},{\bf r}_0;\omega_0) \cdot {\bf d}_{\rm 
ge}\right|^{2}\, .
\end{equation}

In the limit $k_0 |{\bf r}- {\bf r}_0| \gg 1$, the Green function $\mathbb{G}({\bf r},{\bf r}_0;\omega_0)$ can be evaluated from 
Eq.~\eqref{GreenFunction} using the stationary phase method (where the fast oscillating phase is given by $\bm{k}\cdot\bm{x} + i k_z^{(s)} |z|$). 
The obtained result is
\begin{eqnarray}
\label{Assymptotic_FF}
\!\!\!\!\!\!\!\!\mathbb{G}({\bf r},{\bf r}_{0};\omega_0)\simeq \dfrac{-ik_{n}|z|}{2\pi 
r^{2}}e^{ik_{n}\left(\!r - \frac{z^2}{r}\!\right)}\tilde{\mathbb{G}}\!\left(\dfrac{{\bm x}}{r}k_{n},z, d;\omega_0\right),
\end{eqnarray}
with $k_n=k_0$ for $z>0$ and $k_n=k_0\sqrt{\varepsilon_{s}/\varepsilon_0}$ for $z<0$.
Using this result together with equations (\ref{ReflectedGreenFunction}), (\ref{TransmittedGreenFunction}), (\ref{AveragePowerDef}), and (\ref{FF_Poynting}), one can put the total power emitted into the far field by the dipole as \cite{Novotny}
\begin{equation}
 P_{\rm rad} = P_{\rm rad}^{\rm up} +  P_{\rm rad}^{\rm down}\, ,
\end{equation}
where $P_{\rm rad}^{\rm up}$ and $P_{\rm rad}^{\rm down}$ are the average powers emitted into the regions $z>0$ and $z<0$, respectively. Splitting each 
term into contributions from dipole components that are perpendicular and parallel to the $XY$-plane, performing the integral over $\phi$ and changing the variable of integration $\theta$ according to $k = k_0 \sin \theta$ for $P_{\rm rad}^{\rm up}$ and $k = k_0 \sqrt{\varepsilon_s/\varepsilon_0} \sin \theta$ for $P_{\rm 
rad}^{\rm down}$, we can express the average powers as
\begin{widetext}
\begin{align}
\label{radiative_up_perp}
\frac{P_{{\rm rad},\perp}^{{\rm up}}}{P_{0}} & 
= \frac{1}{2}+\frac{3}{4}\int_{0}^{k_{0}}\!\!\!\!dk\frac{k^3}{k_{0}^3\left|k_{z}\right|}\left\{ 2\text{Re}\left[ e^{2ik_{z}d}r^{{\rm 
p,p}}\right] +\left|r^{{\rm p,p}}\right|^{2}+\left|r^{{\rm s,p}}\right|^{2}\right\},\\
\label{radiative_up_par}
\frac{P_{{\rm rad},\parallel}^{{\rm up}}}{P_{0}} & = \frac{1}{2}+\frac{3}{8}\int_{0}^{k_{0}}\!\!\!\!dk \frac{k}{k_{0}\left|k_{z}\right|}\left\{ 
2\text{Re}\left[e^{2ik_{z}d}\left(r^{{\rm s,s}}-\frac{\left|k_{z}\right|^{2}}{k_{0}^{2}}r^{{\rm p,p}}\right)\right]+
 \left|r^{{\rm s,s}}\right|^{2}+\frac{\left|k_{z}\right|^{2}}{k_{0}^{2}}\left|r^{{\rm p,p}}\right|^{2}+\left|r^{{\rm 
p,s}}\right|^{2}+\frac{\left|k_{z}\right|^{2}}{k_{0}^{2}}\left|r^{{\rm s,p}}\right|^{2}\right\}, \\
\label{radiative_down_perp}
\frac{P_{{\rm rad},\perp}^{{\rm down}}}{P_{0}} & = 
\frac{3}{4}\int_{0}^{k_0 \sqrt{\varepsilon_s/\varepsilon_0}}\!\!\!\!dk\frac{k^3}{ k_{0}^3 \left|k_{z}\right| 
}e^{-2d\text{Im}k_{z}}\frac{\left|k_{z}^{
s}\right|}{\left|k_{z}\right|}\left\{\left|t^{{\rm p,p}}\right|^{2}+\left|t^{{\rm s,p}}\right|^{2}\right\},\\
\label{radiative_down_par}
\frac{P_{{\rm rad},\parallel}^{{\rm down}}}{P_{0}} & = 
\frac{3}{8}\int_{0}^{k_0 \sqrt{\varepsilon_s/\varepsilon_0}}\!\!\!\!dk \frac{k}{ k_{0} \left|k_{z}\right| }e^{-2d\text{Im}k_{z}}\frac{\left|k_{z}^{
s}\right|}{\left|k_{z}\right|}\left\{\left|t^{{\rm s,s}}\right|^{2}+\frac{\left|k_{z}\right|^{2}}{k_{0}^{2}}\left|t^{{\rm 
p,p}}\right|^{2}+\left|t^{{\rm p,s}}\right|^{2}+\frac{\left|k_{z}\right|^{2}}{k_{0}^{2}}\left|t^{{\rm s,p}}\right|^{2}\right\},
\end{align}
\end{widetext}
where $P_0 = \omega_0^{4} |{\bf d}_{\rm ge}|^2 /(3 \pi \varepsilon_0 c^3) =
\hbar \omega_0 \Gamma_0$ is the total power emitted in free space. For the power
emitted into the $z<0$ region, there are two different contributions: $(i)$
$\pi/2 < \theta < \arcsin ( \sqrt{\varepsilon_s / \varepsilon_0}
)$, or $k_0<k<k_0\sqrt{\varepsilon_s / \varepsilon_0}$, which is
usually referred to as {\it forbidden light region} \cite{Novotny} and
corresponds to an inverted total internal reflection process, in
which a decaying wave that is emitted by the dipole is transmitted as a
propagating wave once it reaches the interface (we refer to this contribution as
$P_{\rm rad}^{\rm down, f}$); $(ii)$ $ \arcsin (\sqrt{\varepsilon_s
/ \varepsilon_0}) < \theta < \pi$, or $k<k_0$, which corresponds to the
emission of propagating waves (we refer to this contribution as $P_{\rm 
rad}^{\rm down, a}$). By subtracting the power emitted via 
the radiative processes ($P_{\rm rad}$) from the total dissipated power ($P_{\rm total} = \hbar 
\omega_0 \Gamma$), we obtain the power dissipated via non-radiative processes ($P_{\rm non-rad}$). The contributions of the perpendicular em parallel components of the electric dipole to 
non-radiative power can be written  in terms of  absorption  coefficients as
\begin{eqnarray}
\!\!\!\!\!\!\!\!\dfrac{P_{{\rm non-rad},\perp}}{P_{0}} & =& \frac{3}{4}\int_{0}^{+\infty}\!\!\!\!\!\!dk\dfrac{k^{3}e^{-2d\text{Im}k_{z}}}{\left|k_{z}\right|k_{0}^{3}}\mathcal{A}_{\rm p}
\label{non-radiative_perp}\, ,\\
\!\!\!\!\!\!\!\!\dfrac{P_{{\rm non-rad},\parallel}}{P_{0}} & =& 
\dfrac{3}{8}\int_{0}^{+\infty}\!\!\!\!\!\!dk\dfrac{k e^{-2d\text{Im}k_{z}}}{\left|k_{z}\right|k_{0}}\left(\mathcal{A}_{\rm 
s}+\dfrac{\left|k_{z}\right|^{2}}{k_{0}^{2}} 
\mathcal{A}_{\rm p}\right),
\label{non-radiative_parallel}
\end{eqnarray}
where the absorption coefficients are given by
\begin{widetext}
\begin{equation}
\label{absorption_p}
 \mathcal{A}_{\rm p}=\begin{cases}
1-\left[\left|r^{{\rm p,p}}\right|^{2}+\left|r^{{\rm s,p}}\right|^{2}+\dfrac{\left|k_{z}^{s}\right|}{\left|k_{z}\right|}\left(\left|t^{{\rm 
p,p}}\right|^{2}+\left|t^{{\rm s,p}}\right|^{2}\right)\right] & ,\, k<k_{0}\\
2\text{Im}\left[r^{{\rm p,p}}\right]-\dfrac{\left|k_{z}^{s}\right|}{\left|k_{z}\right|}\left(\left|t^{{\rm p,p}}\right|^{2}+\left|t^{{\rm 
s,p}}\right|^{2}\right) & ,\, k_{0}<k<k_0 \sqrt{\varepsilon_s/\varepsilon_0}\\
2\text{Im}\left[r^{{\rm p,p}}\right] & ,\, k_0 \sqrt{\varepsilon_s/\varepsilon_0}<k
\end{cases}
\end{equation}
\end{widetext}
and $\mathcal{A}_{\rm s}$ is obtained from Eq. (\ref{absorption_p}) by swapping $s\leftrightarrow p$.

\begin{figure}[h!]
\centering
\includegraphics[scale=0.4]{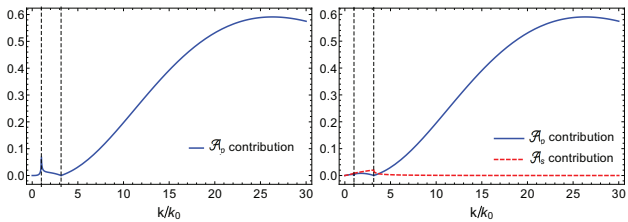}
\caption{Integrands of Eqs.~\eqref{non-radiative_perp} and \eqref{non-radiative_parallel} for the quantum emitter's dissipated power by 
non-radiative processes. On the left we plot the the integrand of Eq.~\eqref{non-radiative_perp}, on the right we plot the integrand of
\eqref{non-radiative_parallel}, splitting it into the individual contributions from the absorption coefficients $\mathcal{A}_{\rm p}$ and 
$\mathcal{A}_{\rm s}$. The vertical dashed lines mark the points $k=k_0$ and $k=k_0 \sqrt{\varepsilon_s / \varepsilon_0}$. The values of $B=5$ T and 
$\mu_c=115$ meV were used.}
\label{Fig5}
\end{figure}

Note that for 
$k>k_0\sqrt{\varepsilon_s / \varepsilon_0}$ the expressions for the non-radiative emitted power, Eqs.~\eqref{non-radiative_perp} and 
~\eqref{non-radiative_parallel}, coincide with the expressions for the total SE rate, Eqs.~\eqref{SEPerp} and \eqref{SEPar}. Hence, we can 
interpret the integration region $k>k_0\sqrt{\varepsilon_s / \varepsilon_0}$ in Eqs.~\eqref{SEPerp} and \eqref{SEPar} as being a contribution to the SE rate exclusively due to non-radiative processes. 
The region $k<k_0\sqrt{\varepsilon_s / 
\varepsilon_0}$ also contributes to the non-radiative decay, as can be seen in Fig.~\ref{Fig5}, where the integrands of 
Eqs.~\eqref{non-radiative_perp} and \eqref{non-radiative_parallel} as a function of $k$ are plotted. This is only expected as propagating waves emitted 
by the dipole can also be absorbed and dissipated by the graphene layer. It should be mentioned, however, that the contribution of wavevectors $k<k_0\sqrt{\varepsilon_s / \varepsilon_0}$ to the non-radiative SE decay is negligible  when compared to the contribution coming from $k>k_0\sqrt{\varepsilon_s / \varepsilon_0}$ (see Fig. \ref{Fig5}). Therefore, the non-radiative decay due to LSW can be well approximated  by Eq.~\eqref{non_radiative}. In the  same way, we can approximate the contribution to the SE rate from $k<k_0\sqrt{\varepsilon_s / \varepsilon_0}$ in Eqs.~\eqref{SEPerp} and 
\eqref{SEPar} as being exclusively owing to radiative processes.  As such we can approximate $P_{{\rm rad}, \perp}^{\rm down, f}$ (TIR modes) by 
Eq.~\eqref{radiative_forbidden} and $P_{{\rm rad}, \perp }^{\rm down, a} + P_{{\rm rad}, \perp }^{\rm up}$ (propagating modes) by Eq.~\eqref{radiative_allowed}. 
Approximations \eqref{radiative_allowed}-\eqref{radiative_forbidden} were tested numerically against the exact results and the differences were found 
to be negligible. 

Finally, we notice that the power emitted by the quantum emitter that is absorbed by graphene due to Joule heating can be written as
\begin{eqnarray}
\!\!\!\!\!\!\!\!\!\!\!\! P_{{\rm g}}&=&2\mu^{2}\omega_{0}^{4}\ \text{Re}\!\!\int\dfrac{d^{2}\bm{k}}{\left(2\pi\right)^{2}}\cr 
 && \!\!\!\!\!\!\!\!\!\!\!\! \times{\bf d}_{{\rm 
ge}}^{*}\!\cdot\!\tilde{\mathbb{G}}^{\dagger}\left(\bm{k},0,0;\omega_{0}\right)\!\cdot\!\bm{\sigma}\left(\bm{k},\omega_{0}\right)\!\cdot\!\tilde{\mathbb{G}}
\left(\bm { k } , 0 ,
0;\omega_{0}\right)\!\cdot\!{\bf d}_{ge}\, .
\end{eqnarray}
 In the case when $\text{Im} [\varepsilon_s]=0$, 
non-radiative decay is exclusively due to graphene and it is possible to show that the power absorbed by graphene can be written as 
Eqs.~\eqref{non-radiative_perp} and \eqref{non-radiative_parallel}, with absorption coefficients given by Eq.~\eqref{absorption_p}.

\begin{thebibliography}{99}
\bibitem{note3} It must be noticed that the conductivity that enters the boundary conditions is an one particle 
irreducible quantity (with respect to the EM propagator), as solving the Maxwell's equations automatically resumes bubble diagrams, 
including screening effects.
\end{thebibliography}
\end{document}